\documentclass[aps,pra]{revtex4}
\usepackage{amsmath}

\begin{document}

\title{Alternative implementation of atomic form factors}

\author{Abdaljalel Alizzi}
\email{abdaljalel90@gmail.com}\affiliation{Novosibirsk State University, 
Novosibirsk 630 090, Russia}

\author{Abhijit Sen}
\email{abhijit913@gmail.com}\affiliation{Novosibirsk State University, 
Novosibirsk 630 090, Russia}

\author{Z.~K.~Silagadze}
\email{Z.K.Silagadze@inp.nsk.su}\affiliation{Budker Institute of
Nuclear Physics and Novosibirsk State University, Novosibirsk 630
090, Russia }

\begin{abstract}
Using a new result on the integral involving the product of Bessel 
functions and associated Laguerre polynomials, published in the mathematical 
literature some time ago, we present an alternative method for calculating
discrete-discrete transition form factors for hydrogen-like atoms. For 
comparison, an overview is also given of two other commonly used methods.
\end{abstract}

\maketitle

\section{introduction}
A bound state of muon and anti-muon (true muonium or dimuonium), although
predicted long ago \cite{1,2,3,4}, has never been observed experimentally.
Many mechanisms for the production of dimuonium have been proposed in the 
literature. Dimuonium can be formed in fixed-target experiments \cite{5,6,
7,8,9}, in electron-positron collisions \cite{1,10,11,12}, in elementary 
particle decays \cite{13,14,15,16,17,18,19}, in a quark-gluon plasma 
\cite{20,21}, in relativistic heavy ion collisions \cite{21,22,23}, in an 
astrophysical context \cite{24}, or in experiments with ultra-slow muon beams
\cite{25,26}.

As part of the first stage of the expensive and long-term super charm-tau 
factory project, Budker Institute of Nuclear Physics (Novosibirsk) is 
currently developing plans to build an inexpensive, low-energy 
$\mu\mu$-tron machine \cite{27}.  Apart from purely accelerator studies, the 
$\mu\mu$-tron will make it possible to produce and investigate dimuonium 
experimentally. Studying the interactions of dimuonium with ordinary atoms 
as it passes through the foil is an integral part of the planned experiments.

Elementary atoms, such as dimuonium, when passing through the foil interact 
with ordinary atoms predominantly via the Coulomb potential \cite{28A,28,29}. 
Such an interaction is treated in terms of atomic form factors, and a
comprehensive review of atomic form factor calculations can be found in 
\cite{30}. 

General analytical formulas for calculating the form factor of a hydrogen-like 
atom were obtained in \cite{31} by group-theoretical methods. However, these
formulas have a somewhat complicated form, requiring time-consuming 
calculations for each value of a transfer momentum \cite{32}.

A more convenient set of formulas was developed in \cite{32,33} and  
implemented as a FORTRAN program in \cite{34}. Based on the mathematical 
results obtained in \cite{35}, in this article we present an alternative 
method for calculating the form factor, which in some sense complements 
the method presented in \cite{32,33}. 

As a byproduct of this reaearch, some trigonometric identities involving 
Chebyshev polynomials of the second kind were obtained in \cite{36}.

Throughout the paper, we use dimuonium atomic units, in which $c=\hbar=1$,
the unit of mass is $\frac{1}{2}m_\mu$ (reduced mass in the dimuonium atom), 
and the unit of length is the radius of the first Bohr orbit in dimuonium:
$a_B=2\,\frac{m_e}{m_\mu}\,a_0\approx 512~\mathrm{fm}$, where $a_0\approx
5.29\times 10^{-11}~\mathrm{m}$ is the usual Bohr radius.

Although the motivation for the article was Novosibirsk dimuonium program, 
we emphasize that the results obtained are in fact of much broader interest, 
mainly in atomic physics, see Devangan's review article \cite{30}.
 
\section{Cross sections of dimuonium interactions with an external field}
The differential cross section for the scattering of a particle with mass 
$m_i$ and initial momentum $p_i = |\vec {p} _i|$ on the potential 
corresponding to the interaction Hamiltonian $ H_ {int}$ is given by the 
expression \cite{37}
\begin{equation}
\frac{d\sigma}{d\Omega}=(2\pi)^4\frac{m_im_fp_f}{p_i}|T_{fi}|^2,
\label{eq1}
\end{equation}
where $m_f$, $p_f=|\vec{p}_f|$ are the mass and momentum of the particle after 
scattering, and in the Born approximation
\begin{equation}
T_{fi}=\langle\Psi_f|H_{int}|\Psi_i\rangle.
\label{eq2}
\end{equation}
Let us apply this relations in the case of scattering of dimuonium in the 
Coulomb field of the target atomic nucleus. If $\vec{r}_1$ is the radius 
vector of a muon in a dimuonium atom, and $\vec{r}_2$ is the radius vector of 
an anti-muon, then the wave function of dimuonium with momentum $\vec{P}$ and
the center of mass radius vector $\vec{R}=(m_1\vec{r}_1+m_2\vec{r}_2)/ 
(m_1+m_2)$ is given by
\begin{equation}
\Psi(\vec{r}_1,\vec{r}_2)=(2\pi)^{-3/2}e^{i\vec{P}\cdot\vec{R}}\varphi(\vec{r}),
\label{eq3}
\end{equation}
where $m_1=m_2=m_\mu$, $\vec{r}=\vec{r}_1-\vec{r}_2$, and $\varphi(\vec{r})$ is 
the Coulomb wave function of the relative motions of muon and anti-muon in 
the dimuonium atom. If $U(\vec{r})$ is the (screened) Coulomb field of the 
target nucleus, then
\begin{equation}
T_{fi}=\int d\vec{r}_1\,d\vec{r}_2\,\Psi^*_f(\vec{r}_1,\vec{r}_2)\left [
eU(\vec{r}_1)-eU(\vec{r}_2)\right ]\Psi_i(\vec{r}_1,\vec{r}_2),
\label{eq4}
\end{equation}
where $e$ is the muon charge.

It can be checked that changing of variables from $\vec{r}_1$, $\vec{r}_2$ to
$\vec{R}$, $\vec{r}$ in the multiple integral (\ref{eq4}) has a unit Jacobian.
Using (\ref{eq3}) and
\begin{equation}
U(\vec{r})=\frac{1}{(2\pi)^3}\int d\vec{q}\,e^{i\vec{q}\cdot\vec{r}}\,\tilde{U}
(\vec{q}),\;\;\vec{r}_1=\vec{R}+\frac{1}{2}\,\vec{r},\;\;
\vec{r}_1=\vec{R}-\frac{1}{2}\,\vec{r},
\label{eq5}
\end{equation}
in (\ref{eq4}), we get after a simple integration in $d\vec{R}$ (which 
produces $\delta(\vec{q}+\vec{P}_i-\vec{P}_f)$ $\delta$-function)
\begin{equation}
T_{fi}=\frac{e\,\tilde{U}(\vec{q})}{(2\pi)^3}\,\left [F_{fi}\left(\frac{\vec{q}}
{2}\right )-F_{fi}\left(-\frac{\vec{q}}{2}\right )\right ],
\label{eq6}
\end{equation}
where
\begin{equation}
F_{fi}(\vec{q})=\int d\vec{r}\,\varphi_f^*(\vec{r})\,e^{i\vec{q}\cdot\vec{r}}\,
\varphi_i(\vec{r}),
\label{eq7}
\end{equation}
and $\vec{q}=\vec{P}_f-\vec{P}_i$. Therefore, according to (\ref{eq1}),
\begin{equation}
\frac{d\sigma}{d\Omega}=\frac{1}{(2\pi)^2}\,\frac{M^2P_f}{P_i}\,e^2\,
\left|\tilde{U}(\vec{q})\right|^2\left |F_{fi}\left(\frac{\vec{q}}
{2}\right )-F_{fi}\left(-\frac{\vec{q}}{2}\right )\right |^2,
\label{eq8}
\end{equation}
where $M\approx 2m_\mu$ is the dimuonium mass.

If $\theta$ is dimuonium scattering angle, then $q^2=P_f^2+P_i^2-2P_fP_i
\cos{\theta}$ and $qdq=P_fP_i\sin{\theta}d\theta$, which imply
\begin{equation}
d\Omega=2\pi\sin{\theta}d\theta=2\pi\,\frac{qdq}{P_iP_f}.
\label{eq9}
\end{equation}
Besides, if the initial $\varphi_i$ and final $\varphi_f$ quantum states have 
definite angular momenta $l$ and $l^\prime$, then
\begin{equation}
F_{fi}(-\vec{q})=\int d\vec{r}\,\varphi_f^*(\vec{r})\,e^{-i\vec{q}\cdot\vec{r}}\,
\varphi_i(\vec{r})=\int d\vec{r}\,\varphi_f^*(-\vec{r})\,e^{i\vec{q}\cdot\vec{r}}\,
\varphi_i(-\vec{r})=(-1)^{l+l^\prime}F_{fi}(\vec{q}).
\label{eq10}
\end{equation}
In light of (\ref{eq8}), (\ref{eq9}) and (\ref{eq10}), we can write the 
$(n,l,m)\to (n^\prime,l^\prime,m^\prime)$  discret-discret transition cross 
section in the scattering of dimuonium by the target nucleus electric 
field $U$ in the form
\begin{equation}
d\sigma_{nlm}^{n^\prime l^\prime m^\prime}=\frac{e^2\left(1-(-1)^{l-l^\prime}\right)}
{\pi V^2}\,\left|\tilde{U}(\vec{q})\right|^2\left |F_{nlm}^{n^\prime l^\prime m^\prime}
\left(\frac{\vec{q}}{2}\right )\right |^2 qdq,
\label{eq11}
\end{equation}
where $V=P_i/M$ is the initial velocity of dimuonium, and we used the 
fact that $(-1)^{l+l^\prime}=(-1)^{l-l^\prime+2l^\prime}=(-1)^{l-l^\prime}$, and, therefore
$\left(1-(-1)^{l+l^\prime}\right)^2=1-2(-1)^{l-l^\prime}+(-1)^{2(l-l^\prime)}=
2\left(1-(-1)^{l-l^\prime}\right)$.

Summation over the complete set of final states gives the following sum rule:
\begin{equation}
\sum\limits_f \left |F_{fi}\left(\frac{\vec{q}}{2}\right )-F_{fi}\left(-
\frac{\vec{q}}{2}\right )\right |^2=2\left (1-F_{ii}(\vec{q}\,)\right ).
\label{eq12}
\end{equation}
Indeed, using $\sum\limits_f |f\rangle\langle f|=1$, we get 
\begin{eqnarray} &&
\sum\limits_f \left |F_{fi}\left(\frac{\vec{q}}{2}\right )-F_{fi}\left(-
\frac{\vec{q}}{2}\right )\right |^2= \sum\limits_f \left\langle i\left|\left (
e^{-i\frac{\vec{q}\cdot\vec{r}}{2}}-e^{i\frac{\vec{q}\cdot\vec{r}}{2}}\right )\right|f\right
\rangle\left\langle f\left | \left ( e^{i\frac{\vec{q}\cdot\vec{r}}{2}}-
e^{-i\frac{\vec{q}\cdot\vec{r}}{2}}\right )\right |i\right\rangle = \nonumber \\ && 
\left\langle i\left |\left (e^{-i\frac{\vec{q}\cdot\vec{r}}{2}}-
e^{i\frac{\vec{q}\cdot\vec{r}}{2}}\right )\left ( e^{i\frac{\vec{q}\cdot\vec{r}}{2}}-
e^{-i\frac{\vec{q}\cdot\vec{r}}{2}}\right )\right|i\right\rangle=2\langle i|i\rangle
-\langle i|(e^{i\vec{q}\cdot\vec{r}}+e^{-i\vec{q}\cdot\vec{r}})|i\rangle=2\left (1-
\mathrm{Re}\,F_{ii}(\vec{q}\,)\right).
\label{eq13}
\end{eqnarray}
But, if the initial $\varphi_i$ quantum state has a definite angular momentum
$l$, then (\ref{eq7}) and (\ref{eq10}) show that $F_{ii}(\vec{q}\,)$ is real:
\begin{equation}
F^*_{ii}(\vec{q}\,)=F_{ii}(-\vec{q}\,)=(-1)^{2l}F_{ii}(\vec{q}\,)=F_{ii}(\vec{q}\,),
\label{eq14}
\end{equation}
and (\ref{eq12}) does follow.

With the help of the sum rule (\ref{eq12}), we can calculate the total cross 
section of dimiunium transitions from the initial $(n,l,m)$ quantum state
to some final state (discrete or continuum) in the following way:
\begin{equation}
d\sigma_{nlm}^{tot}=\frac{e^2}{\pi V^2}\,\left|\tilde{U}(\vec{q})\right|^2
\left [1-F_{nlm}^{nlm}(\vec{q}\,)\right ] qdq.
\label{eq15}
\end{equation}
The main results of this section, equations (\ref{eq11}) and (\ref{eq15}),
were obtained long ago in \cite{28A}. Here we present their derivations for
the sake of reference and to establish the notation.

\section{Master formula for form factor calculation}  
As equations (\ref{eq11}) and (\ref{eq15}) show, the central object in the
study of interactions of dimuonium with matter is the hydrogen-like 
discrete-discrete atomic form-factor (which is just the Fourier transform of 
$\varphi^*_{n_2l_2m_2}(\vec{r}\,)\,\varphi_{n_1l_1m_1}(\vec{r}\,)$ with respect to 
the transferred momentum $\vec{q}$ \cite{30})
\begin{equation}
F_{n_1l_1m_1}^{n_2l_2m_2}(\vec{q})=\int d\vec{r}\,\varphi^*_{n_2l_2m_2}(\vec{r}\,)\,
e^{i\vec{q}\cdot\vec{r}}\,\varphi_{n_1l_1m_1}(\vec{r}\,),\;\;\varphi_{n_1l_1m_1}
(\vec{r}\,)=R_{nl}(r)Y_{lm}(\Omega),
\label{eq16}
\end{equation}
where $Y_{lm}$ are usual spherical functions and the hydrogen-like radial wave
functions $R_{nl}(r)$ has the form
\begin{equation}
R_{nl}(r)=\frac{2}{n^2}\sqrt{\frac{(n-l-1)!}{(n+l)!}}e^{-r/n}\left(\frac{2r}
{n}\right)^l L_{n-l-1}^{2l+1}\left(\frac{2r}{n}\right),
\label{eq17}
\end{equation}
with $L_n^m$ as the associated Laguerre polynomials. 

Some words of caution are maybe appropriate here: unfortunately, the 
definitions of neither the ordinary nor the associated Laguerre polynomials 
used in the literature are universal \cite{38}. In the mathematical literature,
two standard definitions of the associated Laguerre polynomials are used:
the definition of Arfken and Weber \cite{39} and the definition of  Spiegel 
\cite{40}. We use the first one, so 
\begin{equation}
L_n^m(x)=(n+m)!\sum\limits_{k=0}^n\frac{(-1)^k}{k!(n-k)!(k+m)!}\,x^k.
\label{eq18}
\end{equation}
In old physics literature, often some variant of the Spiegel's definition
is used. For example, the associated Laguerre polynomials $\tilde{L}_n^m$ used
by Landau and Lifshitz in \cite{41} are $ \tilde{L}_n^m(x)=(-1)^mn!\,L_{n-m}^m(x)$
and they differ  from the  Spiegel's definition by a factor $n!$.

Using plane-wave expansion \cite{37}
\begin{equation}
e^{i\vec{q}\cdot\vec{r}}=4\pi\sum_{l=0}^\infty\sum_{m=-l}^l i^l j_l(qr)
Y_{lm}(\Omega_q)Y^*_{lm}(\Omega_r),
\label{eq19}
\end{equation}
in (\ref{eq16}), we get
\begin{equation}
F_{n_1l_1m_1}^{n_2l_2m_2}=4\pi \sum_{l=0}^\infty\sum_{m=-l}^l \int\limits_0^\infty r^2 dr\,
R^*_{n_2l_2}(r)R_{n_1l_1}(r)i^l j_l(qr)Y_{lm}(\Omega_q)
I_{\;l_1\;\;l_2\;\;l}^{m_1m_2m}\;,
\label{eq20}
\end{equation}
where
\begin{equation}
I_{\;l_1\;\;l_2\;\;l}^{m_1m_2m}=\int d\Omega\, Y^*_{l_2m_2}(\Omega)Y^*_{lm}(\Omega)
Y_{l_1m_1}(\Omega)=(-1)^{m_2+m}\int d\Omega\, Y_{l_2,-m_2}(\Omega)Y_{l,-m}(\Omega)
Y_{l_1m_1}(\Omega).
\label{eq21}
\end{equation}
The angular integral (\ref{eq21}) can be expressed in terms of Wigner's 
$3j$-symbols \cite{41}:
\begin{equation}
I_{\;l_1\;\;l_2\;\;l}^{m_1m_2m}=
(-1)^{m_2+m}\sqrt{\frac{(2l_1+1)(2l_2+1)(2l+1)}{4\pi}}\left(\begin{array}{ccc}
l_1 & l_2 & l \\ 0 & 0 & 0 \end{array}\right )\left(\begin{array}{ccc}
l_1 & l_2 & l \\ m_1 & -m_2 & -m \end{array}\right ).
\label{eq22}
\end{equation}
Energy conservation equation
\begin{equation}
\frac{\vec{P}_i^2}{2M}+\epsilon_i=\frac{\vec{P}_f^2}{2M}+\epsilon_f,
\label{eq23}
\end{equation}
where $\epsilon_i$ and $\epsilon_f$ are dimuonium energy eigenvalues before 
and after scattering in the target nucleus electric field, indicate that
$\vec{P}_f^2\equiv (\vec{P}_i+\vec{q})^2=\vec{P}_i^2+2M(\epsilon_i-\epsilon_f)
\approx \vec{P}_i^2$, that is $\vec{q}\cdot\vec{P}_i\approx 0$ (note that
$\vec{P}_i^2\gg-2M\epsilon_i$ is the condition for the validity of the Born 
approximation \cite{28}). Then, if the quantization axis ($z$-axis) is along 
the initial flow direction (i.e. in the direction of $\vec{P}_i$), 
$\cos{\theta_q}\approx 0$. But $Y_{lm}(\theta,\varphi)\sim P_l^m(\cos{\theta})$ 
and $P_l^m(-\cos{\theta})=(-1)^{l+m}P_l^m(\cos{\theta})$. Therefore, 
$Y_{lm}(\Omega_q)$ is nonzero only if $(-1)^{l+m}=1$.  On the other hand, 
$3j$-symbol  $\left(\begin{array}{ccc}l_1 & l_2 & l \\ 0 & 0 & 0 \end{array}
\right )$ is nonzero only if $l+l_1+l_2$ is an even integer \cite{42}, and
$\left(\begin{array}{ccc}l_1 & l_2 & l \\ m_1 & -m_2 & -m \end{array}\right )$
is nonzero only if $m=m_1-m_2$ and $|l_1-l_2|\le l\le l_1+l_2$. Therefore, the 
sum (\ref{eq20}) is nonzero only if
$(-1)^{l_1-m_1}=(-1)^{(l_1+l_2+l)-(l+m)-2l_2+l_2-m_2}=(-1)^{l_2-m_2}$.
As we see, in discrete-discrete atomic transitions of dimuonium, the 
$z$-parity  $P_z=(-1)^{l-m}$ is conserved \cite{43}. 

Substituting (\ref{eq17}) into (\ref{eq20}), we get after some algebra
\begin{equation}
F_{n_1l_1m_1}^{n_2l_2m_2}=N \sum\limits_{l=|l_1-l_2|}^{l_1+l_2}
A_l \,I_l=
N\sum\limits_{s=0}^{\mathrm{min}(l_1,\,l_2)}A_{L-2s}\,I_{L-2s},
\label{eq24}
\end{equation}
where
\begin{equation}
N=\frac{(2a)^{l_1+1}\,(2b)^{l_2+1}}{n_1+n_2}\sqrt{(2l_1+1)(2l_2+1)\,
\frac{(n_1-l_1-1)!\,(n_2-l_2-1)!}{(n_1+l_1)!\,(n_2+l_2)!}},
\label{eq25}
\end{equation}

\begin{equation}
A_l=i^l(-1)^{m_2+m}\sqrt{4\pi(2l+1)}\left(\begin{array}{ccc}
l_1 & l_2 & l \\ 0 & 0 & 0 \end{array}\right )\left(\begin{array}{ccc}
l_1 & l_2 & l \\ m_1 & -m_2 & -m \end{array}\right )Y_{lm}(\Omega_q),
\label{eq26}
\end{equation}

\begin{equation}
I_l=\int\limits_0^\infty x^{l_1+l_2+2}\,e^{-x}\,j_l(\sigma x)\,L_{n_1-l_1-1}^{2l_1+1}(2ax)
\,L_{n_2-l_2-1}^{2l_2+1}(2bx)\,dx,
\label{eq27}
\end{equation}
and we have introduced notations \cite{34}
\begin{equation}
a=\frac{n_2}{n_1+n_2},\;\;b=\frac{n_1}{n_1+n_2},\;\;\sigma=\frac{n_1n_2}
{n_1+n_2}\,q,\;\;x=\frac{r}{ab(n_1+n_2)},\;\;s=\frac{1}{2}(L-l),\;\;
L=l_1+l_2.
\label{eq28}
\end{equation}
At last, if the quantization axis is parallel to the collision direction,
we have \cite{34} 
\begin{equation}
Y_{lm}(\Omega_q)\approx Y_{lm}\left (\frac{\pi}{2},\varphi\right )=
(-1)^m2^{-l}e^{im\varphi}\cos{\left(\frac{\pi}{2}(l+m)\right )}\frac{\sqrt{(l+m)!\,
(l-m)!}}{\Gamma\left(1+\frac{l+m}{2}\right)\,\Gamma\left(1+\frac{l-m}{2}
\right)}\sqrt{\frac{2l+1}{4\pi}},
\label{eq29}
\end{equation}
since, as we have seen, the transferred momentum $\vec{q}$ is almost 
perpendicular to this axis (which is parallel to $\vec{P}_i$). 

A much simpler expression
\begin{equation}
Y_{lm}(\Omega_q)=Y_{lm}(0,\varphi)=\delta_{m0}\,\sqrt{\frac{2l+1}{4\pi}},
\label{eq30}
\end{equation}
corresponds to the case when the direction of the transferred momentum 
$\vec{q}$ is taken as the direction of the quantization axis. However,
the first choice is preferable when the initial momentum of the flying 
dimuonium is much larger than the transferred momentum, since in this case 
the quantization axis will remain almost unchanged in successive collisions 
\cite{34}.

\section{Calculation of the radial integral $I_l$ \'{a} la Dewangan}
The most straightforward way to evaluate the integral (\ref{eq27}), outlined
in \cite{30}, is to use the expression (\ref{eq18}) of the associated Laguerre 
polynomials. Applying (a variant of) the Cauchy product formula \cite{44}
\begin{equation}
\sum\limits_{m_1=0}^{M_1}\sum\limits_{m_2=0}^{M_2}a_{m_1m_2}=\sum\limits_{k=0}^{M_1+M_2}
\sum\limits_{m_1=0}^k a_{m_1,k-m_1},
\label{eq31}
\end{equation}
we get
\begin{equation}
L_{n_1-l_1-1}^{2l_1+1}(2ax)\,L_{n_2-l_2-1}^{2l_2+1}(2bx)=\sum\limits_{k=0}^{n_1+n_2-L-2}
C_k\,x^k,
\label{eq32}
\end{equation}
where 
\begin{equation}
C_k=\sum\limits_{j=0}^k\frac{(-1)^k(n_1+l_1)!\,(n_2+l_2)!\,(2a)^j(2b)^{k-j}}
{j!\,(k-j)!\,(n_1-l_1-1-j)!\,(n_2-l_2-1+j-k)!\,(2l_1+1+j)!\,(2l_2+1+k-j)!}.
\label{eq33}
\end{equation}
Substitution of this result into (\ref{eq27}) yields
\begin{equation}
I_l=\sum\limits_{k=0}^{n_1+n_2-L-2}C_k\,\tilde{J}_k,
\label{eq34}
\end{equation}
where
\begin{equation}
\tilde{J}_k=\int\limits_0^\infty x^{L+k+2}\,e^{-x}\,j_l(\sigma x)\,dx=
\sqrt{\frac{\pi}{2\sigma}}\int\limits_0^\infty x^{L+k+\frac{3}{2}}\,e^{-x}\,
J_{l+\frac{1}{2}}(\sigma x)\,dx.
\label{eq35}
\end{equation}
But $L+k+\frac{3}{2}=2s+k+1+l+\frac{1}{2}$ and  (\ref{eq35}) can be rewritten 
as follows 
\begin{equation}
\left . \tilde{J}_k=\sqrt{\frac{\pi}{2\sigma}}\left(-\frac{\partial}{\partial
\alpha}\right)^{2s+k+1}\int\limits_0^\infty x^{l+\frac{1}{2}}\,e^{-\alpha x}\,
J_{l+\frac{1}{2}}(\sigma x)\,dx\right |_{\alpha=1}.
\label{eq36}
\end{equation}
The integral entering in this expression can be found in the classical table 
of integrals by Gradshteyn and Ryzhik \cite{45} (entry 6.623.1. In \cite{46} 
this integral is simply evaluated by the heuristic method of brackets):
\begin{equation}
\int\limits_0^\infty e^{-\alpha x}J_\nu(\beta x)\,x^\nu\,dx=\frac{(2\beta)^\nu
\Gamma\left(\nu+\frac{1}{2}\right)}{\sqrt{\pi}\left(\alpha^2+\beta^2\right)^
{\nu+\frac{1}{2}}}.
\label{eq37}
\end{equation}
Besides we have \cite{30}
\begin{equation}
\left(\frac{\partial}{\partial\alpha}\right)^s\frac{1}{[\alpha^2+
\sigma^2]^{l+1}}=\sum\limits_{p=0}^{[s/2]}\frac{(-1)^{s+p}\,s!\,(l+s-p)!\,\,
(2\alpha)^{s-2p}}{(s-2p)!\,p!\,l!\,[\alpha^2+\sigma^2]^{l+1+s-p}},
\label{eq38}
\end{equation}
where $[s/2]$ denotes the integer part of $s/2$ (the largest integer
$\le s/2$). In light of (\ref{eq37}) and (\ref{eq38}), we finally get
\begin{equation}
\tilde{J}_k=(2\sigma)^{L-2s}\sum\limits_{p=0}^{s+[\frac{k+1}{2}]}\frac{(-1)^p
(2s+k+1)!\,(L-p+k+1)!\,2^{2(s-p)+k+1}}{[2(s-p)+k+1]!\,p!\,(1+\sigma^2)^{L-p+k+2}}.
\label{eq39}
\end{equation}
Equations (\ref{eq34}), (\ref{eq33}), (\ref{eq34}) and (\ref{eq39}) determine
the atomic form factor $F_{n_1l_1m_1}^{n_2l_2m_2}$ as a four-fold finite series
of rational functions of $q$.

\section{Calculation of  the radial integral $I_l$ \'{a} la Afanasyev and 
Tarasov}
Now we outline calculation of $I_l$ as given in \cite{33}. The starting point
will be a Clebsch-Gordan-type linearisation relation for the product of two
associated Laguerre polynomials obtained in \cite{47} (valid for $a+b=1$):
\begin{equation}
L_n^\alpha(ax)\,L_m^\beta(bx)=\sum\limits_{k=0}^{n+m}C_{nm}^{\alpha\beta}(a,b)
L_k^{\alpha+\beta}(x),
\label{eq40}
\end{equation}
where
\begin{equation}
C_{nm}^{\alpha\beta}(a,b)=\frac{k!\,(n+m-k)!\,}{n!\,m!}\,a^{k-m}\,b^{k-n}\,
P_{n+m-k}^{(k-m,\,k-n)}(b-a)\,P_{n+m-k}^{(\alpha+k-m,\,\beta+k-n)}(b-a),
\label{eq41}
\end{equation}
and $P_n^{(\alpha,\beta)}(x)$ are Jacobi polynomials. Therefore,
\begin{equation}
L_{n_1-l_1-1}^{2l_1+1}(2ax)\,L_{n_2-l_2-1}^{2l_2+1}(2ax)=\sum\limits_{k=0}^{n_1+n_2-L-2}
H_k\,L_k^{2(L+1)}(2x),
\label{eq42}
\end{equation}
with
\begin{equation}
H_k=C_{n_1-l_1-1,\,n_2-l_2-1}^{2l_1+1,\,2l+2+1}(a,b).
\label{eq43}
\end{equation}
Nota that Afanasyev and Tarasov don't cite \cite{47} and provide their own
derivation of (\ref{eq42}) in \cite{33} with seemingly different result for
$H_k$. However, it can be shown that their result is equivalent to 
(\ref{eq43}) since
\begin{equation}
P_n^{(-m,\,-k)}(x)=\left(\frac{x-1}{2}\right)^m\left(\frac{x+1}{2}\right)^k
P_{n-m-k}^{(m\,k)}(x).
\label{eq44}
\end{equation}
This last identity can be proved by using the relation
\begin{equation}
P_n^{(-m,\beta)}(x)=\frac{\Gamma(n+\beta+1)}{\Gamma(n+\beta+1-m)}
\frac{(n-m)!}{n!}\left(\frac{x-1}{2}\right)^m\,P_{n-m}^{(m,\beta)}(x),
\label{eq45}
\end{equation}
which can be found in the book \cite{48} (formula 4.22.2), in combination with 
the symmetry relation 
\begin{equation}
P_n^{(\alpha,\beta)}(x)=(-1)^n\,P_n^{(\beta,\alpha)}(-x).
\label{eq46}
\end{equation}
Next, using the formula \cite{49}
\begin{equation}
\sum_{k=0}^n\frac{(-1)^k}{(n-k)!}\binom{m-n}{k}\left(\frac{2}{z}\right)^k
J_{m-k}(z)=\frac{(-1)^n}{n!}J_{m-2n}(z),\;\;\;m>n,
\label{eq47}
\end{equation}
continued analytically for a non-integer $m=L+\frac{1}{2}$ and rewritten in 
terms of spherical Bessel functions $j_n(z)=\sqrt{\frac{\pi}{2z}}\,J_{n+1/2}(z)$, 
Afanasyev and Tarasov expand  
\begin{equation}
j_{L-2s}(\sigma x)=\sum_{p=0}^sB_{ps}\left(\frac{2}{\sigma x}\right)^p
j_{L-p}(\sigma x),
\label{eq48}
\end{equation}
with
\begin{equation}
B_{ps}=(-1)^{s-p}\,\Gamma(p+1)\binom{s}{p}\binom{L-s+1/2}{p}.
\label{eq49}
\end{equation}
In light of (\ref{eq42}) and (\ref{eq48}), (\ref{eq27}) takes the form
\begin{equation}
I_l=\sum\limits_{p=0}^s\sum\limits_{k=0}^{n_1+n+2-L-2}B_{ps}\,H_k\,\left(\frac{2}
{\sigma}\right)^pI_k^{(L,\,p)}(\sigma),
\label{eq50}
\end{equation}
where
\begin{equation}
I_{\;k}^{(L,\,p)}(\sigma)=\int\limits_0^\infty x^{L-p+2}e^{-x}j_{L-p}(\sigma x)\,
L_k^{2L+2}(2x)\,dx.
\label{eq51}
\end{equation}
Generating function for Associated Laguerre polynomials \cite{39}
\begin{equation}
\frac{e^{-tx/(1-t)}}{(1-t)^{m+1}}=\sum\limits_{n=0}^\infty L_n^m(x)\,t^n
\label{eq52}
\end{equation}
indicates that
\begin{equation}
 L_n^m(x)=\left . \frac{1}{n!}\,\frac{\partial^n}{\partial t^n}\,
\frac{e^{-tx/(1-t)}}{(1-t)^{m+1}}\right |_{t=0}.
\label{eq53}
\end{equation}
Therefore, we can rewritten (\ref{eq51}) in the following form
\begin{equation}
I_{\;k}^{(L,\,p)}(\sigma)=\left .\frac{1}{k!}\,\frac{\partial^k}{\partial t^k}\,
\left (\frac{1}{(1-t)^{2L+3}}\int\limits_0^\infty x^{L-p+2}\,e^{-\frac{1+t}{1-t}\,x}\,
j_{L-p}(\sigma x)\,dx\right )\right |_{t=0}.
\label{eq54}
\end{equation}
Now it is clear why the expansion (\ref{eq48}) was needed: it simplifies
the calculation of the integral in (\ref{eq54}) through (\ref{eq37}).
Indeed, with $\alpha=\frac{1+t}{1-t}$ we have
\begin{eqnarray} && 
\int\limits_0^\infty x^{L-p+2}\,e^{-\alpha\,x}\,j_{L-p}(\sigma x)\,dx=
\sqrt{\frac{\pi}{2\sigma}}\int\limits_0^\infty x^{L-p+\frac{3}{2}}\,
e^{-\alpha\,x}\,J_{L-p+\frac{1}{2}}(\sigma x)\,dx= \nonumber \\ &&
\sqrt{\frac{\pi}{2\sigma}}
\left(-\frac{\partial}{\partial \alpha}\right)\int\limits_0^\infty 
x^{L-p+\frac{1}{2}}\,e^{-\alpha\,x}\,J_{L-p+\frac{1}{2}}(\sigma x)\,dx=
\frac{2\alpha\,(2\sigma)^{L-p}\,(L-p+1)!}{(\alpha^2+\sigma^2)^{L-p+2}},
\label{eq55}
\end{eqnarray}
where at the last step we have used (\ref{eq37}). Therefore, we get after 
some elementary algebra
\begin{equation}
I_{\;k}^{(L,\,p)}(\sigma)=\frac{2\,(2\sigma)^{L-p}\,(L-p+1)!}{(1+\sigma^2)^{L-p+2}}\,
\left .\frac{1}{k!}\,\frac{\partial^k}{\partial t^k}\left (\frac{1+t}
{(1-t)^{2p}\left(1+t^2-2\,\frac{\sigma^2-1}{\sigma^2+1}\,t\right)^{L-p+2}}
\right )\right |_{t=0}.
\label{eq56}
\end{equation}
If $k\ge 1$, using Leibnitz's formula for the k-th derivative of the product 
of two functions, we get
\begin{equation}
\left .\frac{1}{k!}\,\frac{\partial^k}{\partial t^k}\left (t\,f(t)\right )
\right |_{t=0}=\left .\frac{1}{(k-1)!}\,\frac{\partial^{k-1}}{\partial t^{k-1}}
\left (f(t)\right )\right |_{t=0},
\label{eq57}
\end{equation}
and finally
\begin{equation}
I_{\;k}^{(L,\,p)}(\sigma)=\frac{2\,(2\sigma)^{L-p}\,(L-p+1)!}{(1+\sigma^2)^{L-p+2}}\,
\left [C_{\;k}^{(L+2,p)}\left(\frac{\sigma^2-1}{\sigma^2+1}\right)+
C_{\;k-1}^{(L+2,p)}\left(\frac{\sigma^2-1}{\sigma^2+1}\right)\right ],
\label{eq58}
\end{equation}
where
\begin{equation}
C_{\;k}^{(\lambda,\,p)}(x)=\left .\frac{1}{k!}\,\frac{\partial^k}{\partial t^k}
\left (\frac{1}{(1-t)^{2p}(1+t^2-2tx)^{\lambda-p}}\right)\right |_{t=0}
\label{eq59}
\end{equation}
It is clear from (\ref{eq56}) that (\ref{eq58}) will be valid even for $k=0$
if we take $C_{\;-1}^{(\lambda,\,p)}(x)=0$.

Using
\begin{equation}
\frac{1}{(1-t)^{2p}}=\sum\limits_{n=0}^\infty\binom{n+2p-1}{n}t^n,\;\;
\frac{1}{(1+t^2-2tx)^{\lambda-p}}=\sum\limits_{n=0}^\infty C_{\;n}^{\lambda-p}(x)\,t^n,
\label{eq60}
\end{equation}
where $C_{\;n}^{(\alpha)}(x)$ are Gegenbauer (ultraspherical) polynomials 
\cite{39}, and the Cauchy product formula \cite{44}, we get
\begin{equation}
f(t)=\frac{1}{(1-t)^{2p}(1+t^2-2tx)^{\lambda-p}}=\sum\limits_{k=0}^\infty\left [
\sum\limits_{l=0}^k\binom{l+2p-1}{l}C_{\;k-l}^{(\lambda-p)}(x)\right ]t^k.
\label{eq61}
\end{equation} 
Therefore
\begin{equation}
C_{\;k}^{(\lambda,\,p)}(x)=\sum\limits_{l=0}^k\binom{l+2p-1}{l}C_{\;k-l}^{(\lambda-p)}(x)
\label{eq62}
\end{equation}
can be considered as generalized Gegenbauer polynomials. It is convenient
to calculate them by using the recurrence relation, which we will now derive.

Note that
\begin{equation}
\frac{\frac{\partial}{\partial t} f(t)}{f(t)}=\frac{\partial}{\partial t}\ln{
f(t)}=\frac{2p}{1-t}+\frac{2(\lambda-p)(x-t)}{1+t^2-2tx}=\frac{2\left[
\lambda t^2+\left (p(1-x)-\lambda(1+x)\right)t+p(1-x)+\lambda x\right]}
{-t^3+(1+2x)t^2-(1+2x)t+1}\equiv \frac{g_1(t)}{g_2(t)}.
\label{eq63}
\end{equation}
On the other hand, since
\begin{equation}
g_1(0)=2[p(1-x)+\lambda x],\;\;\left .\frac{\partial g_1}{\partial t}
\right |_{t=0}=2[p(1-x)-\lambda(1+x)],\;\; \left .\frac{\partial^2 g_1}{\partial 
t^2}\right |_{t=0}=4\lambda,
\label{eq64}
\end{equation}
and
\begin{equation}
g_2(0)=1,\;\;\left .\frac{\partial g_2}{\partial t}\right |_{t=0}=-(1+2x),\;\; 
\left .\frac{\partial^2 g_2}{\partial t^2}\right |_{t=0}=2(1+2x),\;\;
\left .\frac{\partial^3 g_2}{\partial t^3}\right |_{t=0}=-6,
\label{eq65}
\end{equation}
we will get
\begin{eqnarray} &&
\left . \frac{1}{(k+1)!}\,\frac{\partial^{k+1}}{\partial t^{k+1}}\left (g_2(t)
\frac{\partial f(t)}{\partial t}\right )\right |_{t=0}=\left .
\sum\limits_{n=0}^{k+1}\binom{k+1}{n}\frac{\partial^n g_2(t)}{\partial t^n}\,
\frac{1}{(k+1)!}\frac{\partial^{k+2-n} f(t)}{\partial t^{k+2-n}}\right |_{t=0}=
\nonumber \\ &&
(k+2)C_{k+2}^{(\lambda,\,p)}(x)-(k+1)(1+2x)C_{k+1}^{(\lambda,\,p)}(x)+k(1+2x)
C_k^{(\lambda,\,p)}(x)-(k-1)C_{k-1}^{(\lambda,\,p)}(x),
\label{eq66}
\end{eqnarray}
and
\begin{eqnarray} &&
\left . \frac{1}{(k+1)!}\,\frac{\partial^{k+1}}{\partial t^{k+1}}\left (g_1(t)
f(t)\right )\right |_{t=0}=\left .\sum\limits_{n=0}^{k+1}\binom{k+1}{n}
\frac{\partial^n g_1(t)}{\partial t^n}\,
\frac{1}{(k+1)!}\frac{\partial^{k+1-n} f(t)}{\partial t^{k+1-n}}\right |_{t=0}=
\nonumber \\ &&
2[p(1-x)+\lambda x]C_{k+1}^{(\lambda,\,p)}(x)+2[p(1-x)-\lambda(1+x)]
C_k^{(\lambda,\,p)}(x)+k(1+2x)+2\lambda C_{k-1}^{(\lambda,\,p)}(x).
\label{eq67}
\end{eqnarray}
Since acording to (\ref{eq63}) $g_2(t)\frac{\partial f(t)}{\partial t}=
g_1(t)f(t)$, these two expressions must be equal, and we obtain the following
recurrence relation \cite{33}
\begin{eqnarray} 
& (k+2)C_{k+2}^{(\lambda,\,p)}(x)= & \nonumber \\ 
& \left[k+1+2p+2x(k+\lambda-p+1)\right]C_{k+1}^{(\lambda,\,p)}(x)- 
\left[k+2\lambda-2p+2x(k+\lambda+p)\right ]
C_k^{(\lambda,\,p)}(x)+(k+2\lambda-1)C_{k-1}^{(\lambda,\,p)}(x). \quad &
\label{eq68}
\end{eqnarray}
Futher, by using $C_{\;0}^{(\alpha)}(x)=1$, $C_{\;1}^{(\alpha)}(x)=2\alpha x$,
$C_{\;2}^{(\alpha)}(x)=-\alpha+2\alpha(1+\alpha)x^2$, we get from (\ref{eq62})
\begin{equation}
C_{\;0}^{(\lambda,\,p)}(x)=1,\;\;C_{\;1}^{(\lambda,\,p)}(x)=2\left[p+(\lambda-p)x\right],
\;\; C_{\;2}^{(\lambda,\,p)}(x)=2(\lambda-p)(1+\lambda-p)x^2+4p(\lambda-p)x+
2p(1+p)-\lambda.
\label{eq69}
\end{equation}
This gives the initial values for the recurrence relation (\ref{eq68}).

The expressions (\ref{eq50}) and  (\ref{eq58}) allow to calculate the  
radial integral $I_l$ and hence the atomic form factor in numerically more
efficient way compared to the method of the previous section.  

\section{An alternative method for calculating the radial integral $I_l$}
The following mathematical result \cite{35}
\begin{eqnarray} &&
\int\limits_0^\infty e^{-\delta x}J_\nu(\mu x)x^\gamma L_n^\alpha(\beta x)dx=
\sum\limits_{k=0}^n\frac{(-\beta)^k\mu^\nu\Gamma(n+\alpha+1)\Gamma(\nu+
\gamma+k+1)}{k!\Gamma(n-k+1)\Gamma(\alpha+k+1)2^\nu\Gamma(\nu+1)
\delta^{\nu+\gamma+k+1}}
\times \nonumber \\ &&
{_2F_1}\left(\frac{\nu+\gamma+k+1}{2},\frac{\nu+
\gamma+k+2}{2};1+\nu;-\frac{\mu^2}{\delta^2}\right).
\label{eq70}
\end{eqnarray}
can be used to envisage an alternative way for calculating the radial integral
$I_l$. Note that this result was not available to the authors of \cite{33}, 
since \cite{35} was published much later.

According to (\ref{eq18}),
\begin{equation}
L_{n_2-l_2-1}^{2l_2+1}(2bx)=(n_2+l_2)!\sum\limits_{k=0}^{n_2-l_2-1}\frac{(-1)^k(2bx)^k}
{k!\,(n-2-l_2-1-k)!\,(2l_2+1+k)!}.
\label{eq71}
\end{equation}
Therefore (\ref{eq27}) can be rewritten as follows
\begin{equation}
I_l=(n_2+l_2)!\sum\limits_{m_2=0}^{n_2-l_2-1}\frac{(-1)^{m_2}(2b)^{m_2}}
{m_2!\,(n_2-l_2-1-m_2)!\,(2l_2+1+m_2)!}\,J_{l,m_2},
\label{eq72}
\end{equation}
where
\begin{equation}
J_{l,m_2}=\int\limits_0^\infty x^{l_1+l_2+2+m_2}\,e^{-x}\,j_l(\sigma x)\,
L_{n_1-l_1-1}^{2l_1+1}(2ax)\,dx=\sqrt{\frac{\pi}{2\sigma}}
\int\limits_0^\infty x^{l_1+l_2+m_2+\frac{3}{2}}\,e^{-x}\,J_{l+\frac{1}{2}}(\sigma x)\,
L_{n_1-l_1-1}^{2l_1+1}(2ax)\,dx.
\label{eq73}
\end{equation}
The integral is of the type (\ref{eq70}), and we get
\begin{eqnarray} &&
J_{l,m_2}=\frac{4^ll!}{(2l+1)!}\sum\limits_{m_1=0}^{n_1-l_1-1}\frac{(-2a)^{m_1}
(n_1+l_1)!\,(l+l_1+l_2+m_1+m_2+2)!\,\sigma^l}{m_1!\,(n_1-l_1-1-m_1)!\,
(2l_1+1+m_1)!\,2^l}\times \nonumber \\ &&
{_2F_1}\left(\frac{l+l_1+l_2+m_1+m_2+3}{2},\,\frac{l+l_1+l_2+m_1+m_2+4}{2};\,
l+\frac{3}{2};\,-\sigma^2\right),
\label{eq74}
\end{eqnarray}
Therefore,
\begin{eqnarray} &&
I_l=\frac{2^ll!}{(2l+1)!}\sum\limits_{m_1=0}^{n_1-l_1-1}\sum\limits_{m_2=0}^{n_2-l_2-1}
\frac{(-1)^{m_1+m_2}(2a)^{m_1}(2b)^{m_2}(n_1+l_1)!\,(n_2+l_2)!\,
(l+l_1+l_2+m_1+m_2+2)!}{m_1!\,m_2!\,(n_1-l_1-1-m_1)!\,(n_2-l_2-1-m_2)!\,
(2l_1+1+m_1)!\,(2l_2+1+m_2)!}\times \nonumber \\ &&
{_2F_1}\left(\frac{l+l_1+l_2+m_1+m_2+3}{2},\,\frac{l+l_1+l_2+m_1+m_2+4}{2};\,
l+\frac{3}{2};\,-\sigma^2\right).
\label{eq75}
\end{eqnarray}
Gauss hypergeometric function ${_2F_1}$ in this formula can be expressed 
in terms of Jacobi polynomials. In particular,
\begin{eqnarray} &&
{_2F_1}\left(\frac{l+l_1+l_2+m_1+m_2+3}{2},\,\frac{l+l_1+l_2+m_1+m_2+4}{2};\,
l+\frac{3}{2};\,-\sigma^2\right)= \nonumber \\ &&
\left \{\begin{array}{l} \left (\cos{\phi}\right )^{2(l+M+2)}\;
\frac{P_{\;M}^{\left (l+\frac{1}{2},\,\frac{1}{2}\right )}(\cos{2\phi})}
{P_{\;M}^{\left (l+\frac{1}{2},\,\frac{1}{2}\right )}(1)},\;\; \mathrm{if}\;\;
l_1+l_2+m_1+m_2-l=2M, \\ \\ \left (\cos{\phi}\right )^{2(l+M+2)}\;
\frac{P_{\;M+1}^{\left (l+\frac{1}{2},\,-\frac{1}{2}\right )}(\cos{2\phi})}
{P_{\;M+1}^{\left (l+\frac{1}{2},\,-\frac{1}{2}\right )}(1)},\;\; \mathrm{if}\;\;
l_1+l_2+m_1+m_2-l=2M+1,\end{array} \right .
\label{eq76}
\end{eqnarray}
where $M$ is an integer, and the angle $\phi$ is defined by $\tan{\phi}=
\sigma$.

To prove (\ref{eq76}), first consider the case $l_1+l_2+m_1+m_2-l=2M+1$.
Using the Pfaff transformation \cite{50}
\begin{equation}
{_2F_1}(a,b;c;x)=(1-x)^{-a}\;{_2F_1}\left (a,c-b;c;\frac{x}{x-1}\right ),
\label{eq77}
\end{equation}
we get
\begin{equation}
{_2F_1}\left (l+M+2,l+M+\frac{5}{2};l+\frac{3}{2};-\tan^2{\phi}\right )=
\left (\cos{\phi}\right )^{2(l+M+2)}\;{_2F_1}\left (l+M+2,-(M+1);l+\frac{3}{2};
\sin^2{\phi}\right ),
\label{eq78}
\end{equation}
and the second expression of (\ref{eq76}) follows taking into account the 
relation between the Jacobi polynomials and the hypergeometric function 
\cite{50}
\begin{equation}
P_{\;n}^{(\alpha,\,\beta)}(x)=\binom{n+\alpha}{b}\;{_2F_1}\left (-n,n+\alpha+
\beta+1;\alpha+1;\frac{1-x}{2}\right )=
P_{\;n}^{(\alpha,\,\beta)}(1)\;{_2F_1}\left (-n,n+\alpha+
\beta+1;\alpha+1;\frac{1-x}{2}\right ),
\label{eq79}
\end{equation}
and the symmetry property of the hypergeometric function ${_2F_1}(a,b;c;x)=
{_2F_1}(b,a;c;x)$.

If $l_1+l_2+m_1+m_2-l=2M$, we can write the hypergeometric function in
(\ref{eq76}) in the form
\begin{equation}
{_2F_1}\left (l+M+2,l+\frac{3}{2}+M;l+\frac{3}{2};-\tan^2{\phi}\right )=
\left (\cos{\phi}\right )^{2(l+M+2)}\;{_2F_1}\left (l+M+2,-M;l+\frac{3}{2};
\sin^2{\phi}\right ),
\label{eq80}
\end{equation}
and from (\ref{eq79}) the first expression of (\ref{eq76}) follows. 

Jacobi polynomials are convenient in that they can be calculated using the 
three-term recurrence relation
\begin{eqnarray} &
P_{\;n+1}^{(\alpha,\,\beta)}(x)=& \nonumber \\ &
\left (\frac{(2n+\alpha+\beta+1)(2n+\alpha+\beta+2)}
{2(n+1)(n+\alpha+\beta+1)}\,x+\frac{(\alpha^2-\beta^2)(2n+\alpha+\beta+1)}
{2(n+1)(n+\alpha+\beta+1)(2n+\alpha+\beta)}\right )P_{\;n}^{(\alpha,\,\beta)}(x)-
\frac{(n+\alpha)(n+\beta)(2n+\alpha+\beta+2)}{(n+1)(n+\alpha+\beta+1)
(2n+\alpha+\beta)}\,P_{\;n-1}^{(\alpha,\,\beta)}(x), &
\label{eq81}
\end{eqnarray}
with
\begin{equation}
P_{\;0}^{(\alpha,\,\beta)}(x)=1,\;\;\; P_{\;1}^{(\alpha,\,\beta)}(x)=\frac{1}{2}\left [
(\alpha+\beta+2)x+\alpha-\beta\right]
\label{eq82}
\end{equation}
as the initial values.

A much simpler expression can be obtained for the diagonal form 
factor $F_{nlm}^{nlm}$, which according to (\ref{eq15}) is required in the 
total cross section calculations, in the case when it is averaged over the 
magnetic quantum number \cite{51}: 
\begin{equation}
F_{nl}(q)=\frac{1}{2l+1}\sum\limits_{m=-l}^lF_{nlm}^{nlm}(q).
\label{eq83}
\end{equation}
Using Uns\"{o}ld identity \cite{52}
\begin{equation}
\sum\limits_{m=-l}^l Y^*_{lm}(\theta,\varphi)\,Y_{lm}(\theta,\varphi)=
\frac{2l+1}{4\pi},
\label{eq84}
\end{equation}
we get from (\ref{eq21})
\begin{equation}
\sum\limits_{m=-l}^l I_{\;l\;\;l\;L}^{mmM}=\frac{2l+1}{\sqrt{4\pi}}\,\delta_{L0}\,
\delta_{M0},
\label{eq85}
\end{equation}
and then (\ref{eq20}) and (\ref{eq83}) indicate that
\begin{equation}
F_{nl}=\int\limits_0^\infty  r^2 dr\,
R^*_{nl}(r)R_{nl}(r) j_0(qr)=\frac{1}{2n}\,\frac{(n-l-1)!}{(n+l)!}\,I_0,
\label{eq86}
\end{equation}
where
\begin{equation}
I_0=\int\limits_0^\infty x^{2l+2}\,e^{-x}\,j_0(\sigma x)\,\left[
L_{n-l-1}^{2l+1}(x)\right]^2\,dx,\;\;\sigma=\frac{qn}{2},\;\; x=\frac{2r}{n}.
\label{eq87}
\end{equation}
Since
\begin{equation}
j_0(\sigma x)=\frac{\sin{\sigma x}}{\sigma x}=\frac{1}{2i\,\sigma x}\,
\left(e^{i\sigma x}-e^{-i\sigma x}\right),
\label{eq88}
\end{equation}
the integral is of the type (see \cite{45}, entry 7.414.4)
\begin{equation}
\int\limits_0^\infty e^{-\beta x}x^\alpha L_n^\alpha(\lambda x)
L_m^\alpha(\mu  x) dx= \frac{\Gamma(m+n+\alpha+1)}{\Gamma(m+1)
\Gamma(n+1)}\,\frac{(\beta-\lambda)^n (\beta-\mu)^m}{\beta^{n+m+\alpha+1}}\,
{_2{F}_1}
\left(-m,-n;-m-n-\alpha;\frac{\beta(\beta-\lambda-\mu)}{(\beta-\mu)
(\beta-\lambda)}\right),
\label{eq89}
\end{equation}
Introducing $\tan{\phi}=\sigma=\frac{nq}{2}$, $\beta=1-i\sigma$, and noting 
that
\begin{equation}
\frac{\beta(\beta-2)}{(\beta-1)^2}=\frac{\beta^*(\beta^*-2)}{(\beta^*-1)^2}
\frac{1+\sigma^2}{\sigma^2}=\frac{1}{\sin^2{\phi}},
\label{eq90}
\end{equation}
we get
\begin{equation}
I_0=\frac{1}{2i\sigma}\,\frac{(2n-1)!}{[(n-l-1)!]^2}\left [
\frac{(\beta-1)^{2(n-l-1)}}{\beta^{2n}}-\frac{(\beta^*-1)^{2(n-l-1)}}{\beta^{*\,2n}}
\right ]
{_2{F}_1}\left(-(n-l-1),-(n-l-1);1-2n;\,\frac{1}{\sin^2{\phi}}\right ).
\label{eq91}
\end{equation}
But $\beta-1=-i\sigma$, $\beta^*-1=i\sigma$, $\beta=1-i\tan{\phi}=
\frac{e^{-i\phi}}{\cos{\phi}}$ and, therefore,
\begin{equation}
\frac{1}{2i\sigma}\left [
\frac{(\beta-1)^{2(n-l-1)}}{\beta^{2n}}-\frac{(\beta^*-1)^{2(n-l-1)}}{\beta^{*\,2n}}
\right ]=(-1)^{n-l-1}\left (\sin{\phi}\right )^{2(n-l)-3}\,\left(\cos{\phi}
\right )^{2l+3}\,\sin{(2n\phi)}.
\label{eq92}
\end{equation}
The hypergeometric function $_2{F}_1$ satisfies the following identity 
\cite{53}:
\begin{equation}
{_2{F}_1}(-n,b;c;z)=\frac{(b)_n}{(c)_n}\,(-z)^n\,{_2{F}_1}\left
(-n,1-c-n;1-b-n;\frac{1}{z}\right ),
\label{eq93}
\end{equation}
where $(x)_n=\Gamma(x+n)/\Gamma(x)$ is the Pochhammer's symbol
with the property 
\begin{equation}
(-x)_n=(-1)^n(x-n+1)_n,
\label{eq94}
\end{equation}
which implies
\begin{equation}
\frac{(-(n-l-1))_{n-l-1}}{(-(2n-1))_{n-l-1}}=\frac{(1)_{n-l-1}}{(n+l+1)_{n-l-1}}=
\frac{(n-l-1)!\,(n+l)!}{(2n-1)!}.
\label{eq95}
\end{equation}
As a result, we get
\begin{eqnarray} &&
{_2{F}_1}\left(-(n-l-1),-(n-l-1);1-2n;\,\frac{1}{\sin^2{\phi}}\right )=
\nonumber \\ &&
\frac{(-1)^{n-l-1}}{\left (\sin{\phi}\right )
^{2(n-l-1)}}\,\frac{(n-l-1)!\,(n+l)!}{(2n-1)!}\,
{_2{F}_1}\left(-(n-l-1),n+l+1);1;\,\sin^2{\phi}\right ).
\label{eq96}
\end{eqnarray}
Combining (\ref{eq86}), (\ref{eq91}), (\ref{eq92}) and (\ref{eq96}), we get
\begin{equation}
F_{nl}=\frac{\sin{(2n\phi)}\,\left (\cos{\phi}\right )^{2l+4}}{n\,\sin{2\phi}}
\,{_2{F}_1}\left(-(n-l-1),n+l+1);1;\,\sin^2{\phi}\right ).
\label{eq97}
\end{equation}
This result was first obtained in \cite{51}. In light of (\ref{eq96}), $F_{nl}$
can be expressed in terms of Jacobi polynomials:
\begin{equation}
F_{nl}=\frac{\sin{(2n\phi)}\,\left (\cos{\phi}\right )^{2l+4}}{n\,\sin{2\phi}}\;
P_{\;n-l-1}^{(0,\,2l+1)}(\cos{2\phi})=\frac{\left (\cos{\phi}\right )^{2l+4}}{n}\,
U_{n-1}(\cos{2\phi})\;P_{\;n-l-1}^{(0,\,2l+1)}(\cos{2\phi}).
\label{eq98}
\end{equation}
Here $U_n(\cos{\phi})=\sin{((n+1)\phi)}/\sin{\phi}$ are Chebyshev polynomials 
of the second kind.

For $l=0$, (\ref{eq98}) further simplifies. The Jacobi polynomials satisfy
the relation \cite{45}
\begin{equation}
P_{\;n}^{(\alpha,\,\beta+1)}(x)=\frac{2}{2n+\alpha+\beta+2}\,\frac{(n+\beta+1)\,
P_{\;n}^{(\alpha,\,\beta)}(x)+(n+1)\,P_{\;n+1}^{(\alpha,\,\beta)}(x)}{1+z}.
\label{eq99}
\end{equation}
Therefore
\begin{equation}
P_{\;n-1}^{(0,\,1)}(\cos{2\phi})=\frac{1}{2\cos^2{\phi}}\left[ P_{n-1}(\cos{2\phi})+
P_n(\cos{2\phi})\right ],
\label{eq100}
\end{equation}
where $P_n(x)=P_n^{(0,\,0)}(x)$ are Legendre polynomials, and (\ref{eq98}) for
$l=0$ takes the form (in agreement with \cite{32})
\begin{equation}
F_{\,n00}^{n00}=F_{n0}=\frac{\cos^2{\phi}}{2n}\,U_{n-1}(\cos{2\phi})
\left[ P_{n-1}(\cos{2\phi})+P_n(\cos{2\phi})\right ].
\label{eq101}
\end{equation}
Since 
\begin{equation}
\cos^2{\phi}=\frac{1}{1+\tan^2{\phi}}=\frac{4}{4+n^2q^2},\;\;\;
\cos{2\phi}=2\cos^2{\phi}-1=\frac{4-n^2q^2}{4+n^2q^2},
\label{eq102}
\end{equation}
the equation (\ref{eq101}) allows to simply compute the diagonal 
atomic form factor $F_{\,n00}^{n00}$ as a rational function of $q^2$ using some 
computer algebra system. Examples are given in the appendix.

\section{Concluding remarks}
The study of the interactions of dimonium with matter requires knowledge of 
atomic form factors. We have presented an overview of two commonly used 
methods for calculating discrete-discrete transition form factors for 
hydrogen-like atoms. An alternative method, described in the previous chapter,
is based on the results of \cite{35} on an integral involving the product of 
Bessel functions and associated Laguerre polynomials. This new method 
complements the methods presented in \cite{30} and \cite{33} in the sense that 
it combines the simplicity and straightforwardness of the first method with 
the computational efficiency of the second.

We have implemented all three methods described in the text in the FORTRAN 
program. As expected, for relatively large quantum numbers ($n\sim 10$), the
first method results in significantly longer runtimes compared to the other 
two. The availability of various computational methods proved to be very
useful at the stage of program development, as it made it possible to 
recognize and correct some subtle errors in the computer program.

The results of \cite{35} have already been used in \cite{54} for analytical 
evaluation of atomic form factors and applied to Rayleigh scattering by 
neutral atoms. However our presentation is more detailed and has very few, 
if any, overlaps with \cite{54}.

In conclusion, we note two important publications that offer other methods for
evaluation of atomic form factors not discussed in this article. In \cite{55} 
the parabolic quantum numbers and the corresponding wave functions were used.
Calculation of the form factor in the parabolic basis is less complicated 
than in the spherical  one. Moreover, if the atom is in a constant electric 
field, the parabolic basis is preferable. However, many applications
require the knowledge of form factors in spherical basis and, correspondingly,
the connection formula between the parabolic and spherical wave functions
should be used to transform the Bersons-Kulsh analytical form factors from
parabolic to spherical basis.

In \cite{57} the so called phase-space distribution method was used to 
calculate the  classical  form  factors for $nlm\to n^\prime l^\prime m$ 
transitions. It was shown that  the classical form factors can be considered
as an effective averaged versions of the their quantum counterparts. 

\section*{Acknowledgments}
The work is supported by the Ministry of Education and Science of the Russian 
Federation and in part by RFBR grant 20-02-00697-a.

\appendix
\section{Diagonal form factors for $nS$ states, $n\le 6$}
\begin{eqnarray} &&
F_{\,100}^{100}=\frac{16}{(4+q^2)^2},\;\;
F_{\,200}^{200}=\frac{1-3\cdot q^2+2\cdot q^4}{(1+q^2)^4}, \;\;
F_{\,300}^{300}=\frac{16(3^9\cdot q^8-6^6\cdot q^6+2^7\cdot3^5\cdot q^4-2^8\cdot3^1
\cdot7^1\cdot q^2+2^8)}
{(4+9q^2)^6},
\nonumber \\ &&
F_{\,400}^{400}=\frac{2^{14}\cdot q^{12}-23\cdot2^{11}\cdot q^{10}+83\cdot2^9\cdot 
q^8-43\cdot5\cdot2^6\cdot q^6+109\cdot2^4\cdot q^4-4\cdot19\cdot q^2+1}
{(1+4q^2)^8},
\nonumber \\ &&
F_{\,500}^{500}=\frac{16}{(4+25q^2)^{10}}\left[5^{17}\cdot q^{16}-2^4\cdot 5^{16}
\cdot q^{14}+37\cdot2^5\cdot3^2\cdot5^{12}\cdot q^{12}-113\cdot10^{10}\cdot q^{10}
+373\cdot2^9\cdot3\cdot5^8\cdot q^8-\right .\nonumber \\ && \left .
29\cdot11\cdot2^{12}\cdot5^6\cdot q^6+271\cdot3\cdot2^{13}\cdot5^3\cdot 
q^4-2^{19}\cdot5^2\cdot q^2+2^{16}\right],
\nonumber \\ &&
F_{\,600}^{600}=\frac{1}{(1+9q^2)^{12}}\left [2\cdot3^{21}\cdot q^{20}-37\cdot5
\cdot3^{18}\cdot q^{18}+73\cdot 3^{19}\cdot q^{16}-859\cdot2^5\cdot3^{13}\cdot 
q^{14}+263\cdot5\cdot2^4\cdot3^{12}\cdot q^{12}-\right .\nonumber \\ && \left .
149\cdot7\cdot2^3\cdot3^{11}\cdot q^{10}+23\cdot17\cdot5\cdot2^3\cdot3^8\cdot 
q^8-1249\cdot2^2\cdot3^6\cdot q^6+383\cdot2\cdot3^4\cdot q^4-29\cdot15
\cdot q^2+1\right ].
\label{A1}
\end{eqnarray}
For $n\le 4$, they agree with the results of \cite{57}.


\begin{thebibliography}{10}
\bibitem{1}
V.~N.~Baier and V.~S.~Synakh, 
Bimuonium production in electron-positron scattering, 
Sov.\ Phys.\ JETP {\bf 14}, 1122 (1962).  
%%CITATION = SPHJA,14,1122;%%

\bibitem{2}
P.~Budini, Reactions with bound states, CERN Tech. Rep. CM-P00056754, 1961.

\bibitem{3}
V.~W.~Hughes and B.~Maglic, True muonium, Bull.\ Am.\ Phys.\ Soc.\ 
{\bf 16}, 65 (1971).

\bibitem{4}
J.~Malenfant, 
Cancellation of the Divergence of the Wave Function at the Origin in Leptonic 
Decay Rates,
Phys.\ Rev.\ D {\bf 36}, 863 (1987).
%%CITATION = doi:10.1103/PhysRevD.36.863;%%

\bibitem{5}
N.~Arteaga-Romero, C.~Carimalo and V.~G.~Serbo,
Production of bound triplet mu+ mu- system in collisions of electrons with 
atoms,
Phys. Rev. A {\bf 62}, 032501 (2000).
%%CITATION = doi:10.1103/PhysRevA.62.032501;%%

\bibitem{6}
E.~Holvik and H.~A.~Olsen,
Creation of Relativistic Fermionium in Collisions of Electrons with Atoms,
Phys. Rev. D {\bf 35}, 2124 (1987).
%%CITATION = doi:10.1103/PhysRevD.35.2124;%%

\bibitem{7}
A.~Banburski and P.~Schuster,
The Production and Discovery of True Muonium in Fixed-Target Experiments,
Phys. Rev. D {\bf 86}, 093007 (2012).
%%CITATION = doi:10.1103/PhysRevD.86.093007;%%

\bibitem{8}
P.~A.~Krachkov and A.~I.~Milstein,
High-energy $\mu^+\mu^-$ electroproduction,
Nucl. Phys. A \textbf{971}, 71 (2018).
%%CITATION = doi:10.1016/j.nuclphysa.2018.01.013;%%

\bibitem{9}
K.~Sakimoto,
Theoretical study of true-muonium $\mu^{+} \mu^{-}$ formation in muon collision 
processes $\mu^{-} + \mu^{+} e^{-}$ and $\mu^{+} + p \mu^{-}$,
Eur. Phys. J. D {\bf 69}, no.12, 276 (2015).
%%CITATION = doi:10.1140/epjd/e2015-60427-6;%%

\bibitem{10}
S.~J.~Brodsky and R.~F.~Lebed,
Production of the Smallest QED Atom: True Muonium ($\mu^+ \mu^-$),
Phys. Rev. Lett. {\bf 102}, 213401 (2009).
%%CITATION = doi:10.1103/PhysRevLett.102.213401;%%

\bibitem{11}
J.~W.~Moffat,
Does a Heavy Positronium Atom Exist?,
Phys. Rev. Lett. {\bf 35}, 1605 (1975).
%%CITATION = doi:10.1103/PhysRevLett.35.1605;%%

\bibitem{12}
S.~M.~Bilenky, V.~Nguyen, L.~L.~Nemenov and F.~G.~Tkebuchava,
%``Production and decay of (muon-plus muon-minus)-atoms,''
Yad. Fiz. {\bf 10}, 812 (1969).
%%CITATION = YAFIA,10,812;%%

\bibitem{13}
Y.~Ji and H.~Lamm,
Scouring meson decays for true muonium,
Phys. Rev. D {\bf 99}, 033008 (2019).
%%CITATION = doi:10.1103/PhysRevD.99.033008;%%

\bibitem{14}
M.~Fael and T.~Mannel,
On the decays $B \to K^{(*)} + $ leptonium,
Nucl. Phys. B {\bf 932}, 370 (2018).
%%CITATION = doi:10.1016/j.nuclphysb.2018.05.015;%%

\bibitem{15}
G.~A.~Kozlov,
On The Problem Of Production Of Relativistic Lepton Bound States In The Decays 
Of Light Mesons,
Sov.\ J.\ Nucl.\ Phys.\  {\bf 48}, 167 (1988).
%%CITATION = SJNCA,48,167;%%

\bibitem{16}
L.~L.~Nemenov,
Atomic decays of elementary particles,
Yad.\ Fiz.\  {\bf 15}, 1047 (1972).
%%CITATION = YAFIA,15,1047;%%

\bibitem{17}
H.~Lamm and Y.~Ji,
Predicting and Discovering True Muonium $(\mu^+ \mu^-)$,
EPJ Web Conf. {\bf 181}, 01016 (2018).
%%CITATION = doi:10.1051/epjconf/201818101016;%%

\bibitem{18}
Y.~Ji and H.~Lamm,
Discovering True Muonium in $K_L\rightarrow (\mu^+\mu^-) \gamma$,
Phys. Rev. D {\bf 98}, 053008 (2018).
%%CITATION = doi:10.1103/PhysRevD.98.053008;%%

\bibitem{19}
X.~Cid Vidal, P.~Ilten, J.~Plews, B.~Shuve and Y.~Soreq,
Discovering True Muonium at LHCb,
Phys. Rev. D {\bf 100}, 053003 (2019).
%%CITATION = doi:10.1103/PhysRevD.100.053003;%%

\bibitem{20}
Y.~Chen and P.~Zhuang,
Dimuonium $(\mu^+\mu^-)$ Production in a Quark-Gluon Plasma,
arXiv:1204.4389 [hep-ph].
%%CITATION = ARXIV:1204.4389;%%

\bibitem{21}
G.~M.~Yu and Y.~D.~Li,
Photoproduction of large transverse momentum dimuonium ($\mu^+ \mu^-$) in 
relativistic heavy ion collisions,
Chin. Phys. Lett. {\bf 30}, 011201 (2013)
%%CITATION = doi:10.1088/0256-307X/30/1/011201;%%

\bibitem{22}
I.~F.~Ginzburg, U.~D.~Jentschura, S.~G.~Karshenboim, F.~Krauss, V.~G.~Serbo 
and G.~Soff,
Production of bound $\mu^+ \mu^-$-systems in relativistic heavy ion collisions,
Phys. Rev. C {\bf 58}, 3565 (1998).
%%CITATION = doi:10.1103/PhysRevC.58.3565;%%

\bibitem{23}
C.~Azevedo, V.~P.~Gon\c{c}alves and B.~D.~Moreira,
True muonium production in ultraperipheral $PbPb$ collisions,
Phys. Rev. C {\bf 101}, 024914 (2020).
%%CITATION = doi:10.1103/PhysRevC.101.024914;%%

\bibitem{24}
S.~C.~Ellis and J.~Bland-Hawthorn,
Astrophysical signatures of leptonium,
Eur. Phys. J. D {\bf 72}, 18 (2018).
%%CITATION = doi:10.1140/epjd/e2017-80488-7;%%

\bibitem{25}
T.~Itahashi, H.~Sakamoto, A.~Sato and K.~Takahisa,
Low Energy Muon Apparatus for True Muonium Production,
JPS Conf. Proc. {\bf 8}, 025004 (2015).
%%CITATION = doi:10.7566/JPSCP.8.025004;%%

\bibitem{26}
K.~Nagamine,
Past, Present and Future of Ultra-Slow Muons,
JPS Conf. Proc. {\bf 2}, 010001 (2014).
%%CITATION = doi:10.7566/JPSCP.2.010001;%%

\bibitem{27}
A.~Bogomyagkov, V.~Druzhinin, E.~Levichev, A.~Milstein and S.~Sinyatkin,
Low-energy electron-positron collider to search and study $\mu^+\mu^-$ bound
state, EPJ Web Conf. {\bf 181}, 01032 (2018).
%%CITATION = doi:10.1051/epjconf/201818101032;%%

\bibitem{28A}
S.~Mr\'{o}wczy\'{n}ski,
Interaction of elementary atoms with matter,
Phys. Rev. A {\bf 33}, 1549 (1986). 
%%CITATION = doi:10.1103/PhysRevA.33.1549;%%

\bibitem{28}
S.~Mr\'{o}wczy\'{n}ski,
Interaction of Relativistic Elementary Atoms With Matter. 1. General Formulas,
Phys. Rev. D {\bf 36}, 1520 (1987).
%%CITATION = doi:10.1103/PhysRevD.36.1520;%%

\bibitem{29}
K.~G.~Denisenko and S.~Mr\'{o}wczy\'{n}ski,
Interaction of Relativistic Elementary Atoms With Matter. 2. Numerical 
Results,
Phys. Rev. D {\bf 36}, 1529 (1987).
%%CITATION = doi:10.1103/PhysRevD.36.1529;%%

\bibitem{30}
D.~P.~Dewangan, 
Asymptotic methods for Rydberg transitions,
Phys. Rep. {\bf 511}, 1 (2012).
%%CITATION = doi:10.1016/j.physrep.2011.10.001;%%

\bibitem{31}
A.~O.~Barut and R.~Wilson,
Analytic group-theoretical form factors of hydrogenlike atoms for discrete and 
continuum transitions,
Phys. Rev. A 40, 1340 (1989).
%%CITATION = doi:10.1103/PhysRevA.40.1340;%%

\bibitem{32}
L.~Afanasyev and A.~Tarasov,  
Elastic form factors of hydrogenlike atoms in n$S$ states,
JINR Preprint E4-93-293 (1993). 
%%CITATION = JINR-E4-93-293;%%

\bibitem{33}
L.~G.~Afanasev and A.~V.~Tarasov,
Breakup of relativistic $\pi^+\pi^-$ atoms in matter,
Phys. Atom. Nucl. {\bf 59}, 2130 (1996).
%%CITATION = PANUE,59,2130;%%

\bibitem{34}
C.~S.~R\'{i}os and J.~S.~Silva, 
An implementation of atomic form factors,
Comp. Phys. Commun. {\bf 151}, 79 (2003).
%%CITATION = doi:10.1016/S0010-4655(02)00687-2;%%

\bibitem{35}
R.~S.~Alassar, H.~A.~Mavromatis and S.~A.~Sofianos,
A New Integral Involving the Product of Bessel Functions and Associated 
Laguerre Polynomials,
Acta Appl. Math. {\bf 100}, 263 (2008).
%%CITATION = doi:10.1007/s10440-007-9183-1;%%

\bibitem{36}
A.~Sen and Z.~K.~Silagadze,
Trigonometric identities inspired by atomic form factor,
Georgian Math. J. {\bf 27}, 441 (2020).
%%CITATION = doi:10.1515/gmj-2019-2083;%%

\bibitem{37}
R.~G.~Newton, {\it Scattering theory of waves and particles} (Springer, New
York, 1982).

\bibitem{38}
C.~E.~Burkhardt and J.~J.~Leventhal, {\it Topics in Atomic Physics} 
(Springer, New York, 2006).

\bibitem{39}
G.~B.~Arfken and H.~J.~Weber, {\it Mathematical Methods for Physicists} 
(Harcourt, New York, 2001).

\bibitem{40}
M.~R.~Spiegel, {\it Mathematical Handbook of Formulas and Tables} (McGraw-Hill, 
NewYork, 1998).

\bibitem{41}
L.~D.~Landau and E.~M.~Lifshitz, {\it Quantum Mechanics (Non-relativistic 
Theory)} (Pergamon Press, Oxford, 1977).

\bibitem{42}
A.~R.~Edmonds, {\it Angular Momentum in Quantum Mechanics} (Princeton 
University Press, Princeton, 1957).

\bibitem{43}
A.~V.~Tarasov and I.~U.~Christova,
The Eikonal theory of interaction of relativistic dimesoatoms with matter atoms,
JINR Communication, P2-91-10, Dubna, 1991.
%%CITATION = JINR-P2-91-10;%%

\bibitem{44}
T.~M.~Apostol, {\it Mathematical Analysis} (Addison-Wesley, Reading, 1974).

\bibitem{45}
I.~S.~Gradshteyn and  I.~M.~Ryzhik, {\it Tables of Integrals, Series and 
Products} (Academic Press, New York, 1994).

\bibitem{46}
I.~Gonzalez, V.~H.~Moll and A.~Straub,
The method of brackets. Part 2: examples and applications, in T.~Amdeberhan, 
L.~Medina  and V.~H.~Moll (Eds.), {\it Gems in Experimental Mathematics}, 
vol. 517 of Contemporary Mathematics, American Mathematical Society, 2010, 
pp. 157--172.
%%CITATION = doi:10.1090/conm/517/10139;%%

\bibitem{47}
A.~W.~Niukkanen,
Clebsch-Gordan-type linearisation relations for the products of Laguerre 
polynomials and hydrogen-like functions,
J. Phys. A: Math. Gen. {\bf 18}, 1399 (1985).
%%CITATION = doi:10.1088/0305-4470/18/9/022;%%

\bibitem{48}
G.~Szeg\"{o}, {\it Orthogonal Polynomials} (American Mathematical Society,
Providence, 1939).

\bibitem{49}
A.~P.~Prudnikov, Yu.~A.~Brychkov and O.~I.~Marichev, {\it Integrals and 
Series. Vol. 2: Special Functions} (Gordon and Breach Science Publishers,
New York, 1992), p. 636.

\bibitem{50}
G.~E.~Andrews, R.~Askey and R.~Roy, {\it Special functions} (Cambridge 
University Press, Cambridge, 1999).

\bibitem{51}
L.~Afanasev, A.~Tarasov and O.~Voskresenskaya,
Total interaction cross-sections of relativistic $\pi^+ \pi^-$ atoms with 
ordinary atoms in the eikonal approach,
J. Phys. G {\bf 25}, B7 (1999).
%%CITATION = doi:10.1088/0954-3899/25/8/701;%%

\bibitem{52}
C.~E.~Burkhardt and J.~J.~Leventhal, {\it Foundations of Quantum Physics} 
(Springer, New York, 2008).

\bibitem{53}
A.~B.~Olde Daalhuis, Hypergeometric function, in {\it NIST handbook of 
mathematical functions} (U.S. Dept. Commerce, Washington, 2010) pp. 383--401.

\bibitem{54}
L.~Safari, J.~P.~Santos, P.~Amaro, K.~J\"{a}nk\"{a}l\"{a} and F.~Fratini,
Analytical evaluation of atomic form factors: Application to Rayleigh 
scattering, J. Math. Phys. {\bf 56}, 052105 (2015).
%%CITATION = doi:10.1063/1.4921227;%%

\bibitem{55}
I.~Bersons and A. Kulsh,
Transition form factor of the hydrogen Rydberg atom,
Phys. Rev. A {\bf 55}, 1674 (1997).
%%CITATION = doi:10.1103/PhysRevA.55.1674;%%

\bibitem{56}
M.~R.~Flannery and D.~Vrinceanu,
Classical and quantal atomic form factors for $nlm\to n^\prime l^\prime m$ 
transitions, Phys. Rev. A {\bf 65}, 022703 (2002).
%%CITATION = doi:10.1103/PhysRevA.65.022703;%%

\bibitem{57}
L.~G.~Afanasyev, Form factors of the $1s$, $2s$, $3s$, and $4s$ states of 
hydrogen-like atoms for discrete transitions, 
Atom. Data Nucl. Data Tabl. {\bf 61}, 31 (1995).
%%CITATION = doi:10.1016/S0092-640X(95)90010-1;%%

\end{thebibliography}
\end{document}